\documentstyle[12pt]{article}
\renewcommand\theequation{\thesection.\arabic{equation}}
\def\ba{\begin{eqnarray}}
\def\ea{\end{eqnarray}}
\def\nn{\nonumber}
\def\S{\Sigma}
\def\M{\cal M}
\def\C{\tilde{\cal C}}
\begin{document}
\setlength{\unitlength}{1mm}
{\hfill  JINR E2-95-28, January 1995 } \vspace*{2cm} \\
\begin{center}
{\Large\bf On the Description of the Riemannian Geometry}
\end{center}
\begin{center}
{\Large\bf in the Presence of Conical Defects}
\end{center}
\bigskip\bigskip\bigskip
\begin{center}
{\bf  Dmitri V.~Fursaev$^{\ast}$ and Sergey N.~Solodukhin$^{\ast\ast}$}
\end{center}
\begin{center}
{\it Bogoliubov Laboratory of Theoretical Physics, \\
Joint Institute for Nuclear Research, \\
141 980, Dubna, Moscow Region, Russia}
\end{center}

\vspace*{2cm}
\begin{abstract}
A consistent approach to the description of integral coordinate invariant
functionals of the metric on manifolds
${\cal M}_{\alpha}$ with conical defects (or singularities) of the topology
$C_{\alpha}\times\Sigma$ is developed.
According to the proposed prescription ${\cal M}_{\alpha}$
are considered as limits of the converging sequences of smooth
spaces. This enables one to give a strict mathematical meaning to a number of
invariant integral quantities on ${\cal M}_{\alpha}$
and make use of them in applications.
In particular, an explicit representation for the Euler numbers
and Hirtzebruch signature in the presence of conical singularities is found.
Also, higher dimensional Lovelock gravity
on ${\cal M}_{\alpha}$ is shown to be well-defined and the gravitational
action in this theory is evaluated. Other series of applications is related
to computation of black hole entropy in the
higher derivative gravity and in quantum 2-dimensional models.
This is
based on its direct statistical-mechanical derivation in the
Gibbons-Hawking approach, generalized to the singular manifolds
${\cal M}_{\alpha}$, and gives the same results as in the other methods.
\end{abstract}

\begin{center}
{\it PACS number(s): 04.60.+n, 12.25.+e, 97.60.Lf, 11.10.Gh}
\end{center}
\vskip 1cm
\noindent $^{\ast}$ e-mail: fursaev@theor.jinrc.dubna.su;
$^{ \ast\ast}$ e-mail: solod@thsun1.jinr.dubna.su

\newpage
\baselineskip=.8cm

\section{Introduction}
\setcounter{equation}0

Thin cosmic strings are known to give rise to remarkable gravitational
effects. They do not affect immediately the local geometry of a space-time
manifold but change instead its global properties. Placing the origin
of the polar coordinate system on the string axis, one reveals a deficit
$2\pi(1-\alpha)$ of the polar angle $\varphi$ \cite{Vilenkin}. Thus, near
the string world sheet $\Sigma$ the space looks like the direct product
$C_{\alpha}\times\Sigma$ where $C_{\alpha}$ is the conical space with the
corresponding ranging of the angle $0\le \varphi \le 2\pi\alpha$.

This peculiarity results in the interesting quantum effects
which have been studied for both simple cones \cite{qft} and around
cosmic strings \cite{strings}. Spaces with analogous features
appear also in the other important physical applications. The well-known
example is the orbifolds occurring in the string compactifications \cite{GSW}.
A similar set of spaces, called conifolds,
has been proposed to generalize the histories included in Euclidean
functional integrals in quantum gravity \cite{11}.
Finally, much attention has been paid recently to conical defects
in connection with thermodynamics in the presence of black-hole \cite{Su},
\cite{SS} and cosmological \cite{FM} horizons,
where the conical angle $\alpha$ is associated with the inverse
temperature of a system.

One should also point out a number of mathematical results. For instance,
the general theory of the Laplace and heat kernel
operators on such a kind of cones has been developed in \cite{15}
and an explicit form of the DeWitt-Schwinger coefficients
has been found out in some cases \cite{kernels}. As well many works were
devoted to the functional determinants and zeta-function on the different types
of orbifolds \cite{orbifolds}.

On the other hand, a consistent
description of the geometrical quantities and invariant functionals
of the metric on the conical defects seems to be absent.
To elucidate this problem, let us remind that a cone is everywhere flat
space (like the plane) except the tip where its curvature $R$ is singular.
Obviously, calculations by means of the standard
formulas of the Riemannian geometry cannot reveal this delta-like singularity,
and other methods must be used to get a correct result.
One of these, based on topological arguments, was suggested many years ago
by Sokolov and Starobinsky \cite{c4} for two-dimensional cones and
used recently in higher dimensions in \cite{c1}. However, an
approach like that does not seem to be quite satisfactory. It
only concerns computation of the scalar curvature,
saying nothing about components of the Riemann tensor, and faces
difficulties under generalization to arbitrary invariant functionals.

In this paper, we consider a more natural recipe how to handle
singularities with a particular topology $C_{\alpha}\times \Sigma$
and use it in the relevant examples.
The corresponding manifolds will be denoted by ${\cal M}_{\alpha}$.
This method is to replace the singular space by a
sequence of regular manifolds. All the integral invariants are then
well-defined, and final results are obtained when the regularization is
taken off. Some aspects of such a regularization have already been discussed
in the literature (see, for instance, \cite{11}, \cite{c5}, \cite{Allen})
and we represent its further development. Thus, we show that although
arbitrary curvature polynomials in such a
procedure turn out to be divergent and depend on regularization,
some specific integral quantities can be finite and have the strict
mathematical meaning. We make use of this fact to derive a number of new
results
of both mathematical and physical interest.

The paper is organized as follows. The regularization method is described
in section 2. Its features are discussed
in detail for two-dimensional cones where the
regularization ambiguity and the structure of the integral of $R^2$ and
Polyakov-Liouville action are investigated.
Then, the technique and results are extended  for the higher dimensional cases.
We evaluate the components of the Riemann tensor on ${\cal M}_{\alpha}$ and
give examples of the functionals being quadratic in the curvature.

A number of consequences and applications is presented in the second
part of the paper, in section 3, which starts with the discussion of the
generalized variational principle on a class of spaces including $
{\cal M}_{\alpha}$.
Then, we analyze the higher order curvature polynomials that can be
defined on ${\cal M}_{\alpha}$.
An important example is the Euler characteristics
and Hirtzebruch signature of ${\cal M}_{\alpha}$ for which
the explicit integral representation is found.
Also the Lovelock gravity turns out to be strictly
defined on the manifolds with conical singularities and we give
the corresponding generalization of the Lovelock action and equations.
Finally, using our technique, we calculate the black hole entropy in the
higher derivative gravity and in quantum 2-dimensional models.
This is based on a direct statistical-mechanical derivation of the entropy
in the Gibbons-Hawking approach generalized to the singular manifolds
and gives the same results as in the other methods.
Some technical moments are clarified in Appendix A.

\bigskip
\section{The method}
\setcounter{equation}0

{\bf A. Two-dimensional cones}
\vskip 0.2cm

{\bf A.1. Integral curvature}
\vskip 0.2cm

The method is worth illustrating when ${\cal M}_{\alpha}$ is a two-dimensional
space with topology of the cone ${\cal C}_{\alpha}$. Then its metric reads
\begin{equation}
ds^2=e^\sigma (d\rho^2+\rho^2d\phi^2)\equiv e^\sigma ds^2_{C}
\label{17}
\end{equation}
where $ds^2_C$ is the line element on ${\cal C}_{\alpha}$,
$\phi$ runs from $0$ to $2\pi\alpha$, and the conformal factor
$\sigma$ is assumed to have the following expansion in the vicinity of
$\rho=0$:
\begin{equation}
\sigma=\sigma_1\rho^2+\sigma_2\rho^4+...~~~.
\label{18}
\end{equation}
In general,  $\sigma_1$ and $\sigma_2$ can be functions of the angle $\phi$
and a possible constant term in (\ref{18}) can be absorbed by redefinition
of $\rho$.

Due to asymptotics (\ref{18}), the singularity comes out only from the
conical metric $ds^2_C$. Hence,
to understand how to introduce the regularization of ${\cal M}_{\alpha}$,
consider an embedding of ${\cal C}_{\alpha}$ in 3-dimensional
(pseudo)Euclidean space.
It can be given by equations
$x=\alpha\rho \cos(\phi/\alpha)$,
$y=\alpha \rho \sin(\phi/\alpha),$ $ z=\sqrt{|1-\alpha^2|} \rho$,
that define the surface
\begin{equation}
z^2-{|1-\alpha^2| \over \alpha^2}(x^2+y^2)=0~~~,~~~z\ge 0~~~.
\end{equation}
Obviously, if $\alpha \neq 1$, there is a singularity at $z=0$ where
one cannot introduce the tangent space and calculate the curvature
in the usual way.

It is easy to "roll off" the cone tip if going from ${\cal C}_{\alpha}$
to a surface ${\tilde {\cal C}}_{\alpha}$ with the equation
$z=\sqrt{|1-\alpha^2|}
f(\rho,a)$ where $f(\rho,a)$ is a smooth function and $a$ is a
regularization parameter such that $\lim_{a\rightarrow 0}f=\rho$.
So far as for ${\cal C}_{\alpha}$ the
function $z$ has a minimum at $\rho=0$, this should also be valid
for $f$ in the case of the regularized surface. Thus, the only additional
condition on $f(\rho,a)$ is $\partial_\rho f|_{\rho=0}=0$ and
the line element on ${\tilde {\cal C}}_{\alpha}$ can be written as
\begin{equation}
ds^2_{{\tilde {\cal C}}}=ud\rho^2+\rho^2d\phi^2~~~,~~~
u=\alpha^2+(1-\alpha^2)(f'_\rho)^2
\label{12}
\end{equation}
where the function $u$ has the following asymptotics
\begin{equation}
u|_{\rho=0}=\alpha^2~~; ~~~u|_{\rho>>a}=1~~~.
\label{13}
\end{equation}
The simplest example of the regularization is that corresponding
to the change of ${\cal C}_{\alpha}$ to a hyperbolic space
\begin{equation}
z^2-{|1-\alpha^2| \over \alpha^2}(x^2+y^2)=a^2~~~,~~~z\ge0~~~,
\label{14}
\end{equation}
\begin{equation}
ds^2_H={{\rho^2 +a^2\alpha^2} \over {\rho^2+a^2}} d\rho^2 +\rho^2 d\phi^2~~~.
\label{15}
\end{equation}

Instead of the singular manifold (\ref{17}) one can use now
the regularized space ${\tilde {\cal M}}_{\alpha}$ with topology of
${\tilde {\cal C}}_{\alpha}$. To proceed, it is convenient to represent
the scalar curvature on it in the form
\begin{equation}
R=e^{-\sigma} R_{\tilde {\cal C}}-e^{-\sigma} \Box_{\tilde {\cal C}} \sigma
\label{20}
\end{equation}
where $R_{\tilde {\cal C}}$ and $\Box_{\tilde {\cal C}}$ are the curvature
and Laplace operator defined with respect to metric (\ref{12}). Then,
by taking into account the form of the regularized volume element
$d\mu=e^\sigma \sqrt{u}\rho d\rho d\phi$ and asymptotics (\ref{13}), one
can evaluate the integral curvature
on ${\tilde {\cal M}}_{\alpha}$
$$
\int_{{\tilde {\cal M}}_{\alpha}} R = 2\pi\alpha
\int_{0}^{\infty}d\rho u'_\rho u^{-{3 \over 2}}
-\int_{0}^{\infty}\int_{0}^{2\pi\alpha}\sqrt{u}\rho d\rho d\phi~
\Box_{\tilde {\cal C}} \sigma=
$$
\begin{equation}
4\pi(1-\alpha)
-\int_{0}^{\infty}\int_{0}^{2\pi\alpha}\sqrt{u}\rho d\rho d\phi~
\Box_{\tilde {\cal C}} \sigma~~~.
\label{16}
\end{equation}
The conical metric results in the first term in the r. h. s. in (\ref{16})
which does not depend on the regularization. The dependence on $u$ appears
only in the second "volume" term in (\ref{16}), but the latter turns out to be
finite in the limit when regularization is taken off and it coincides with
integral curvature computed in the standard way on the smooth domain
${\cal M}_{\alpha}/\Sigma$ of ${\cal M}_{\alpha}$.
(Let us remind that $\Sigma$ denotes the singular set that is the point
$\rho=0$ in the given case.)
Thus, in the limit
$a\rightarrow 0$ one has from
(\ref{16})
\begin{equation}
\lim _{{\tilde {\cal M}}_{\alpha}\rightarrow
{\cal M}_{\alpha}}\int_{{\tilde {\cal M}}_{\alpha}}R =
4\pi(1-\alpha)+
\int_{{\cal M}_{\alpha}/\Sigma}~R~~~.
\label{limR}
\end{equation}
When in the region ${\cal M}_{\alpha}/\Sigma$ the curvature $R$ equals to
zero, equation ({\ref{limR}) reproduces the formula by
Sokolov and Starobinsky \cite{c4}.
This result does not depend on the concrete behaviour of the
regularization function $u$, which can be shown to be a consequence of the
Gauss-Bonnet theorem relating the integral curvature in two dimensions
with the Euler characteristics. A more deep discussion of this point
will be given in section 3.

So far as only singular point $\rho=0$ can give rise to the first term
in (\ref{limR}), one can introduce a local representation for the curvature
on ${\cal M}_{\alpha}$
\begin{equation}
^{(\alpha)}R={2(1-\alpha) \over \alpha} \delta (\rho)~ + R
\label{loc}
\end{equation}
where $\delta (\rho)$ is the delta-function normalized as
$$
\int_{0}^{\infty}\delta (\rho) \rho d\rho=1~~~.
$$

In applications one also
needs to handle with the higher order curvature polynomials or non-local
functionals on conical defects. However, as distinct from the integral
curvature (\ref{limR}), they include in general non-integrable
singularities like $\delta^n (\rho)$.
Let us consider the properties of such functionals on the simplest
examples.
\bigskip

{\bf A.2. Integral of $R^2$}
\vskip 0.2cm

Without loss of generality, assume that metric (\ref{17}) does not depend
on the angle variable $\phi$. In this case, introducing a new radial
coordinate $x={\rho \over a}$, one can on ${\tilde {\cal M}}_{\alpha}$ write
the curvature at the singular point as a decomposition
\begin{equation}
R=(1-a^2\sigma_1 x^2)({1 \over a^2} {u' \over x u^2} -{4\sigma_1 \over u}
+\sigma_1 {x u' \over u^2} + O(a^2))
\end{equation}
where $u' \equiv {\partial u \over \partial x}$.
Using this it can be shown that the following equality
\begin{equation}
\int_{{\tilde {\cal M}}_\alpha}^{}R^2=
\int_{{\cal M}_{\alpha}/\Sigma}^{}R^2  + 2R(0)
I_{1}(\alpha)+X(\alpha,a) ~~~,
\label{R^2}
\end{equation}
\begin{equation}
X(\alpha,a)=
-\frac 14 R(0)I_{2}(\alpha)
+{1 \over a^2} I_3(\alpha)
\label{X}
\end{equation}
holds at small values of the regularization parameter. Here the
relation $-4\sigma_1=R(\rho=0)\equiv R(0)$ has been used
and $I_k(\alpha)$ denote the integrals
\begin{eqnarray}
&&I_1(\alpha)= 2\pi\alpha\int_{0}^{\infty}{dx}({u'_x \over u^{5 \over 2}})
={4\pi \over 3} {(1-\alpha^3) \over \alpha^2}~~~,
\nonumber \\
&&I_2(\alpha)= 2\pi\alpha\int_{0}^{\infty}{dx  x}
({u'_x \over u^2})^2 u^{1 \over 2}~~~,
\nonumber \\
&&I_3(\alpha)= 2\pi\alpha\int_{0}^{\infty}
{dx \over x}({u'_x \over u^2})^2 u^{1 \over 2}~~~.
\label{I}
\end{eqnarray}

The quantity $I_1 (\alpha)$ that enters into the finite part of the integral
$R^2$ does not depend on choice of the regularization function $u(\rho,a)$
and it is determined only by asymptotics (\ref{13}). However,
$I_2(\alpha)$ and $I_3 (\alpha)$ change under variations of
$u(\rho)$, including coordinate transformations.
Taking this into account there is no reason to consider such terms in
(\ref{R^2}) separately and so they were gathered in a combination
$X(\alpha,a)$.
It follows from the form of (\ref{R^2}), (\ref{X}) and (\ref{I})
that $X(\alpha,a)$
should be an invariant function, singular in the limit $a\rightarrow 0$.
It is also important that at small conical deficits
$X(\alpha,a)$ vanishes as fast as $(1-\alpha)^2$ and
only finite terms in (\ref{R^2}) dominate.
This can be proved in general with the help of equation (\ref{12})
but it is better to
demonstrate for particular regularization (\ref{15})
\begin{eqnarray}
&&I_2(\alpha) ={16\pi(1-\alpha^2)^2 \over 15 \alpha^2} {(\alpha^2+3\alpha+1)
\over (1+\alpha)^3}~~~, \nn \\
&&I_3(\alpha) ={8\pi(1-\alpha^2)^2 \over 15 \alpha^4} {(8\alpha^2+9\alpha+3)
\over (1+\alpha)^3}~~~.
\end{eqnarray}

The above consideration teaches that (\ref{R^2}) and other similar invariant
functionals cannot have a strict mathematical meaning in the
presence of conical singularities. Nevertheless,
the structure of
singular terms in these integrals can be described and, as will be shown,
in some important cases all of them cancel each other
or not contribute to the considered quantities leaving there finite terms
not depending on the regularization.
\bigskip

{\bf A.3. Polyakov-Liouville action}
\vskip 0.2cm

This is an example of the non-local
functional playing an important role in two-dimensional quantum gravity
 since it is result of integrating the conformal anomaly. It looks as
follows \cite{Polyakov}
\begin{equation}
W_{PL}=\int R\psi
\label{PL}
\end{equation}
where $\psi$ is a solution of the equation:
\begin{equation}
\Box \psi=R~~~.
\label{psi}
\end{equation}
Consider (\ref{PL})
on the regularized space ${\tilde {\cal M}}_{\alpha}$ and make use of
(\ref{20}).
Then (\ref{psi}) solves as follows
\begin{eqnarray}
&&\psi=-\sigma +\psi_{\tilde{C}}~~~, \nonumber \\
&&\psi_{\C}=-2 \log \rho+ C\int_{}^{\rho}{u^{1 \over 2} \over \rho}d\rho +E
\label{psi1}
\end{eqnarray}
where $\psi_{\C}$ is a  solution of equation (\ref{psi})
for the smoothed cone $\C$ with the curvature
$R_{\C}= u'_{\rho}/(\rho u^{2})$, $C$ and $E$ are constants.
It should be noted that only for $C={2 \over \alpha}$ this can be
written in everywhere regular form
\begin{equation}
\psi_{\C}= {2 \over \alpha}\int_{0}^{\rho}{u^{1 \over 2}-\alpha \over \rho'}
d\rho' + E~~~
\label{newpsi}
\end{equation}
and, moreover,
the function $\psi_{\tilde{\cal C}}$ coincides in the limit $a \rightarrow 0$
 with the corresponding
solution on the conical space $C_{\alpha}$
\begin{equation}
\psi_{\C} \rightarrow \psi_{\cal C}= 2{(1-\alpha) \over \alpha} \log {\rho}
+E~~~.
\label{psi2}
\end{equation}
where the possible singular term $\sim \ln a$ absorbed in redefinition
of constant $E$.
The function $\psi$
is important for analysis of quantum effects on gravitational background
and enters into the formulas for the energy density of the Hawking radiation
and black hole entropy \cite{SS}.

The non-local action (\ref{PL}) on regularized two-dimensional manifold
$\tilde{\cal M}_\alpha$
can be written in a more suitable form
\begin{equation}
W_{PL}[{\tilde {\cal M}}_{\alpha}]=\int_{{\tilde {\cal C}}} \left(
R_{{\tilde {\cal C}}} \psi_{{\tilde{\cal C}}}-2\sigma
R_{{\tilde {\cal C}}}+\sigma\Box_{{\tilde {\cal C}}}\sigma\right)~~~.
\label{PLreg}
\end{equation}
Obviously, the second and third terms in the r. h. s. of (\ref{PLreg})
give a regular contributions when regularization is taken off. Taking
into account equation (\ref{loc}), one gets when $\tilde{C}_{\alpha}
\rightarrow C_{\alpha}$
\begin{equation}
\int_{{\tilde {\cal C}}}
\sigma R_{{\tilde {\cal C}}}~~~\rightarrow~~~4\pi (1-\alpha)\sigma(0)
\label{reg1}
\end{equation}
where $\sigma(0)$ can be zero, if the asymptotics (\ref{18}) is assumed.
Besides, the limit
\begin{equation}
\int_{{\tilde {\cal C}}} \sigma\Box_{{\tilde {\cal C}}}\sigma~~\rightarrow~~
\int_{{\cal C}} \sigma\Box_{{\cal C}}\sigma\equiv
W_{PL}[{\cal M}_{\alpha}/\Sigma]
\label{reg2}
\end{equation}
can be identified with the contribution to the Polyakov-Liouville action
from the regular points of ${\cal M}_{\alpha}$.
The remaining term in (\ref{PLreg}) for $E=0$ has a nonlocal form
\begin{equation}
\int_{{\tilde{\cal C}_{\alpha}}}^{} R_{\tilde{\cal C}}
\psi_{\tilde{\cal C}} = 2\pi\alpha \int_{0}^{\infty}
{u'_\rho \over u^{3 \over 2}} \psi_{\tilde{\cal C}}(\rho) d\rho =
2\pi\alpha\int_{0}^{\infty}{u'_{\rho} \over u^{3/2}}d\rho
\int_{0}^{\rho}{u^{1/2}-\alpha \over u^{3/2}}d\rho'\equiv X(\alpha,a)~~~.
\label{invbox}
\end{equation}
As one can see, $X(\alpha,a)$ is regular in the limit $a\rightarrow 0$
(for the regularization (\ref{14}) dependence on $a$ is absent)
but depends on the form of the regularization function $u$.
{}From equation (\ref{12}) other important property follows that
when $\alpha\rightarrow 1$ the function $X(\alpha,a)$ vanishes as
$(1-\alpha)^2$.

Finally,  one obtains the action (\ref{PL}) on ${\cal M}_{\alpha}$
in the form
\begin{equation}
W_{PL}[{\cal M}_{\alpha}]=W_{PL}[{\cal M}_{\alpha}/\Sigma]+
8\pi(1-\alpha)\psi (\Sigma)+X(\alpha)
\label{PLfin}
\end{equation}
where $\psi=-\sigma$ is a solution of equation (\ref{psi}) when $\alpha=1$,
$\Sigma$ is a singular point and $X(\alpha)\equiv X(\alpha,a=0)$.
Thus the non-local action $W_{PL}$ turns out to be finite (in the limit
$a \rightarrow 0$) but regularization
dependent.

We will return to
equation (\ref{PLfin}) in section 3.

\bigskip
{\bf B. Higher-dimensional case}
\bigskip

The technique can be extended now to higher dimensions.
Let us consider a two-dimensional cone $C_{\alpha}$
embedded in the Riemann $d$-dimensional manifold ${\cal M}_{\alpha}$
so that near the singularity ($\rho=0$) the metric is represented as
\begin{equation}
ds^2=e^{\sigma}(d\rho^2+\rho^2d\phi^2  +\sum_{i,j=1}^{d-2}
(\gamma_{ij}(\theta)+h_{ij}(\theta)\rho^2)d\theta^id\theta^j + ...)
\equiv e^{\sigma}d\tilde{s}^2
\label{8}
\end{equation}
where ... means terms of higher power in $\rho^2$ and
$\phi$ runs from 0 to $2\pi\alpha$. For convenience we prefer to use
the same parametrization as in two dimensions but,
as distinct from this case, the singular set now
is a ($d-2$)--dimensional surface $\Sigma$ with coordinates
$\{\theta^i \}$ and metric $\gamma_{ij}(\theta)$. Near it
${\cal M}_{\alpha}$ looks as a direct product
${\cal C}_{\alpha}\times\Sigma$. One can also consider ${\cal M}_{\alpha}$
having a number of singular sets $\Sigma_i$, each with the corresponding
conical angle $\alpha_i$. Hereafter the metric will be assumed not to depend
on $\varphi$ at least in the small region of $\Sigma$.

The metric (\ref{8}) can be regularized with a parameter $a$ as in two
dimensions by changing the $g_{\rho\rho}$ component in the conical part
\begin{equation}
ds^2=e^{\sigma}( u(\rho,a)d\rho^2+\rho^2d\phi^2  +\sum_{i,j=1}^{d-2}
(\gamma_{ij}(\theta)+h_{ij}(\theta)\rho^2)d\theta^id\theta^j + ...)~~~.
\label{reg}
\end{equation}
The curvature tensors for a manifold ${\tilde {\cal M}}_{\alpha}$ with metric
(\ref{reg}) and evaluation of the geometrical quantities in the limit
$a\rightarrow 0$
are similar to that we considered in two dimensions. Leaving the details for
Appendix A, it should be mentioned that only the two-dimensional
conical part of (\ref{reg}) gives rise to the
singular contributions.

We begin with formulas for components of the Riemann tensor that can be
represented near $\Sigma$ as
\ba
&&^{(\alpha)}R^{\mu\nu}_{\ \ \alpha\beta} = R^{\mu\nu}_{\ \ \alpha\beta}+
2\pi (1-\alpha) \left( (n^\mu n_\alpha)(n^\nu n_\beta)- (n^\mu n_\beta)
(n^\nu n_\alpha) \right) \delta_\Sigma \nn \\
&&^{(\alpha)}R^{\mu}_{ \ \nu} = R^{\mu}_{ \ \nu}+2\pi(1-\alpha)(n^\mu
n_\nu)
\delta_\Sigma \nn                            \\
&&^{(\alpha)}R = R+4\pi(1-\alpha) \delta_\Sigma
\label{curv}
\ea
where $\delta_\Sigma$ is the delta-function: $\int_{\cal M}^{}f \delta_\Sigma=
\int_{\Sigma}^{}f$; $n^k=n^k_\mu dx^\mu$ are two orthonormal vectors
orthogonal to $\Sigma$, $(n_\mu n_\nu)=\sum_{k=1}^{2}n^k_\mu n^k_\nu$
and the quantities $R^{\mu\nu}_{\ \ \alpha\beta}$, $R^{\mu}_{ \ \nu}$ and $R$
are computed in the regular points ${\cal M}_{\alpha}/\Sigma$ by the
standard method.

\bigskip

A consequence of (\ref{curv}) is the following important formula for
the integral curvature of ${\cal M}_\alpha$
\begin{equation}
\int_{{\cal M}_{\alpha}}~^{(\alpha)}R = 4\pi (1-\alpha) A_\Sigma +
\int_{M_{\alpha}/\Sigma}R
\label{intcurv2}
\end{equation}
where $A_{\Sigma}=\int_{\Sigma}$ is the area of
$\Sigma$. Equation (\ref{intcurv2}) already appeared in a number of recent
publications for particular cases \cite{Su} and it was virtually implied in
results of \cite{c1}.  If ${\cal M}_{\alpha}$ has a number of singular
surfaces $\Sigma_i$ with different conical deficits $2\pi(1-\alpha_i)$
then the first term in (\ref{intcurv2}) should be changed
by the sum over all $\Sigma_i$.

\bigskip

As arbitrary functionals on ${\cal M}_{\alpha}$ are concerned,
we give, as an example,
the integrals of quadratic combinations in $R_{\mu\nu\lambda\rho}$.
The chosen regularization leads to the following results
(for details see Appendix A):
\ba
&&\int_{{\cal M}_\alpha}^{}R^2=\int_{{\cal M}_{\alpha}/\Sigma}^{}R^2
+ \left(2I_1-{(5d-8) \over 8(d-1)}I_2\right) \int_{\Sigma}^{}(R-R_\S )+
8\pi(1-\alpha)\int_{\S}^{}R_\S + \nn \\
&&{3 \over 4(d-1)}I_2 \int_{\S}^{}Q_{\Sigma}+Y(\alpha) A_{\Sigma}~~~,
\label{R1}
\ea
\ba
&&\int_{{\cal M}_\alpha}^{}R^2_{\mu\nu}=\int_{{\cal M}_{\alpha}/\Sigma}^{}
R^2_{\mu\nu} + {1 \over (d-1)}\left ({d \over 2}I_1-{(3d -4) \over 16}I_2
\right)
\int_{\S}^{}(R-R_\S )+ \nn \\
&&{1 \over (d-1)} \left(I_1+{1 \over 8}I_2\right)\int_{\S}^{}
Q_{\Sigma}+{1 \over 2}Y(\alpha)A_\S~~~,
\label{Ricci1}
\ea
\ba
&&\int_{{\cal M}_\alpha}^{}R^2_{\mu\nu\alpha\beta}=\int_{{\cal
M}_{\alpha}/\Sigma}^{}
R^2_{\mu\nu\alpha\beta}+ {1 \over (d-1)}\left(2I_1-{d \over 8}I_2\right)
\int_{\S}^{}(R-R_\S )+ \nn \\
&&{1 \over (d-1)}\left(4I_1-{1 \over 4} I_2\right)\int_{\S}^{}
Q_{\Sigma}+Y(\alpha)A_\S
\label{Riemann1}
\ea
where $Y(\alpha)$ is a quantity divergent in the limit $a\rightarrow 0$
and
$$
Q_{\Sigma}=\frac d2 R_{\mu\alpha\nu\beta}n_i^{\mu}n_i^{\nu}n_j^{\alpha}
n_j^{\beta}-R_{\mu\nu}n_i^{\mu}n_j^{\nu}~~~.
$$
For $d=2$ expression (\ref{R1}) coincides with that derived in the
previous section.
As in two dimensions, integrals (\ref{R1})-(\ref{Riemann1}) contain
both divergent ($Y(\alpha)$) and dependent on the regularization
($I_2(\alpha)$) terms and can be brought into the same form as (\ref{R^2}) by
gathering these terms together. In this case the remaining part of (\ref{R1})-
(\ref{Riemann1}) will be a sum of the integrals over the smooth domain of
${\cal M}_{\alpha}$ and regularization-independent additions in the form
of surface integrals depending on either internal or external geometry
of $\Sigma$. Obviously,
one can proceed in this way and obtain similar expressions for functionals
being higher order curvature polynomials on $\M_\alpha$. These examples
follow below.

\bigskip
\section{Applications}
\setcounter{equation}0

\bigskip

{\bf A. Generalized variational principle}
\bigskip

As the first straightforward application of equation (\ref{intcurv2}),
we consider the variational principle generalized
on a class of manifolds admitting conical singularities.
It can be used, for instance, in the description of gravitational effects
caused by cosmic strings.
The gravitational action including a cosmic string with the tension $\mu$
and 2-dimensional worldsheet $\Sigma$ reads
\begin{equation}
W =-{1 \over 16\pi G}\int_{} R + \mu\int_{\Sigma}\equiv
W_{gr} + \mu\int_{\Sigma}~~~.
\label{action1}
\end{equation}
Without loss of generality we assume that manifolds on which (\ref{action1})
is defined do not have the boundaries.
Consider this functional on the spaces ${\cal M}_{\alpha}$ with conical
singularities distributed over $\Sigma$ and represent it,
according to (\ref{intcurv2}), as
follows:
\begin{equation}
W[{\cal M}_{\alpha}] = W_{gr}[{\cal M}_{\alpha}/\Sigma]
+\left(-{(1-\alpha) \over 4\pi G} + \mu\right)\int_{\Sigma}
\label{action2}
\end{equation}
The form of (\ref{action2}) can be used now to find its variations
on the given class of singular spaces, but without fixing the actual
value of the deficit angle at the conical singularity.
Thus, it is easy to see that, apart from the standard Einstein equations
following
from the first regular term in the r. h. s. of (\ref{action2}),
the independent change of the metric on $\Sigma$ results in the additional
condition
\begin{equation}
1-\alpha=4\pi G\mu~~~
\label{eq3}
\end{equation}
being the well-known relation between the string
tension $\mu$ and the conical angle deficit \cite{Vilenkin}.
Condition (\ref{eq3}) is analogous to the  "surface Einstein
equations" in the presence of matter shells \cite{c6} that can also be
obtained from variations of the gravitational functional \cite{c5}.
As is seen, in the absence of strings (\ref{eq3}) is satisfied only at
the vanishing deficit angle, $\alpha=1$. Therefore,
even in the generalized variational principle the extrema of the
Einstein action in vacuum are realized on the smooth manifolds.
The same conclusion was previously derived in \cite{c5} for a minisuperspace
model. On the other hand, spaces with a number of different conical
defects cannot be extrema of the vacuum functional.

\bigskip

{\bf B. Topological characteristics of ${\cal M}_\alpha$}
\bigskip

Let us turn to definition of the Euler numbers $\chi$ and
the Hirtzebruch signature $\tau$ on manifolds with conical singularities.
We are interested in these quantities so far as they are expressed
through the integrals on powers of the Riemann tensor to which
the regularization technique introduced above can be naturally applied.
To be more specific, consider such a characteristic, say $\chi$, on ${\cal M}
_{\alpha}$ as a limit of this quantity taken on the converging sequence
${\tilde {\cal M}}_{\alpha}$
\begin{equation}
\chi[{\cal M}_{\alpha}]\equiv\lim_{{\tilde {\cal M}}_{\alpha}\rightarrow
{\cal M}_{\alpha}} \chi[{\tilde {\cal M}}_{\alpha}]=\chi~~~.
\label{definition}
\end{equation}
By definition, the right hand side of (\ref{definition}) is only
determined by the topology of the smooth spaces and does not depend on the
regularization parameter.
Therefore, topological characteristics like $\chi$ of a singular manifold
${\cal M}_{\alpha}$ simply coincide with those
of ${\tilde {\cal M}}_{\alpha}$ and should be well-defined integral
invariants.
Our aim now is to find a concrete integral representation of $\chi$ and
$\tau$ for ${\cal M}_{\alpha}$.
\bigskip

{\bf B1. Euler numbers}
\vskip 0.2cm

To begin with, let us investigate the simplest
example when ${\cal M}_{\alpha}$ is a closed four-dimensional space with
one singular surface $\Sigma$. For its regularized analog ${\tilde {\cal M}}
_{\alpha}$ the Euler number reads
\begin{equation}
\chi={1 \over 32\pi^2}\int_{{\tilde {\cal
M}}_{\alpha}}^{}(R^2-4R^2_{\mu\nu}+R^2_
{\mu\nu\alpha\beta})~~~.
\label{chi}
\end{equation}
Using (\ref{R1})-(\ref{Riemann1}) and going from ${\tilde {\cal M}}_\alpha$
to ${\cal M}_\alpha$ one obtains from (\ref{chi}) a finite expression
\begin{equation}
\chi [{\cal M}_\alpha]=
{1 \over 32\pi^2}\int_{{\cal M}_{\alpha}/\Sigma}(R^2-4R^2_{\mu\nu}+R^2_
{\mu\nu\alpha\beta})+
(1-\alpha) \chi [\Sigma]~~~,
\label{Euler4}
\end{equation}
where all the terms depending on the regularization are mutually
cancelled. Formula (\ref{Euler4}) gives the desired representation
for $\chi[{\cal M}_{\alpha}]$ in which the first term in the r. h. s.
is the contribution to the integral from the regular points, and
$\chi [\S]={1 \over 4\pi}\int_{\S}^{}R_\S$ is the Euler number of the
surface $\S$.

Equation (\ref{Euler4}) can be generalized to higher even dimensions $d=2p$.
Without loss of generality, we confine ourselves to the compact
spaces without boundaries. In this case, the Euler number of a
$2p$-dimensional smooth manifold ${\cal M}$ is given by the integral
\cite{Euler}
\begin{equation}
\chi= c_p\int_{{\cal M}}^{}{\cal L}_{2p} \sqrt{g}d^{2p}x
\label{EulerD}
\end{equation}
where ${\cal L}_{p}$ is the following quantity
\begin{equation}
{\cal L}_{p}= \epsilon_{\mu_1 \mu_2...\mu_{2p-1}\mu_{2p}}\epsilon^{\nu_1 \nu_2
... \nu_{2p-1}
\nu_{2p}} R^{\mu_1 \mu_2}_{\ \ \nu_1 \nu_2} ...R^{\mu_{2p-1} \mu_{2p}}_{\ \
\nu_{2p-1} \nu_{2p}}
\label{fun}
\end{equation}
and the constant $c_p$ is
\begin{equation}
c_p={1 \over 2^{2(p+1)}\pi^p p!} ~~~.
\label{c_p}
\end{equation}

Now let manifold in (\ref{EulerD}) be a smooth approximation
${\tilde {\cal M}}_{\alpha}$ of a $2p$-dimensional
space $\M_\alpha$ with a singular
$2(p-1)$-dimensional surface $\S$.
Then the Riemann tensor of $\tilde{\M}_{\alpha}$
can be represented as the sum
\begin{equation}
R^{\mu\nu}_{\ \ \alpha\beta}= R^{\mu\nu}_{(reg) \ \alpha\beta}
+R^{\mu\nu}_{(con) \ \alpha\beta}
\label{reg-con}
\end{equation}
of a term remaining regular when the regularization is taken off and
a term $R^{\mu\nu}_{(con) \ \alpha\beta}$
provided by the conical singularity.
The latter has only one non-trivial component (see (\ref{A4})):
\begin{equation}
R^{\phi  \rho}_{(con) \ \phi \rho} ={1 \over a^2} {u'_x \over 2x u^2}
\label{sing}
\end{equation}
where $x={\rho \over a}$ and
it is assumed that ${\tilde{\M}}_\alpha$ in the vicinity of $\S$
is covered by coordinates $\{ \phi, \rho, \theta^i \}$ with metric
(\ref{reg}). Inserting (\ref{reg-con}) into (\ref{fun}) one gets a polynomial
with respect to $R^{\mu\nu}_{(con) \ \alpha\beta}$.
However, due to antisymmetricity
of the $\epsilon$-tensor only the first order of this quantity survives and
\begin{equation}
{\cal L}_{p}= {\cal L}_{p}^{reg}+  4p \epsilon_{\phi \rho i_1...i_{2p-2}}
\epsilon^{\phi  \rho j_1... j_{2p-2}}
R^{\phi \rho}_{(con) \ \phi \rho}
R^{i_1 i_2}_{\ \ j_1 j_2} ...R^{i_{2p-3} i_{2p-2}}_{\ \ j_{2p-3}
j_{2p-2}}
\label{fun1}
\end{equation}
where indices $i_k$ and $j_k$ run from $1$ to $2p-2$. This means that
no singularities appear in (\ref{fun1}) in the limit $a \rightarrow 0$
apart from an integrable $\delta$-function resulting in a surface
addition on $\S$. To evaluate it,
choose the normal vectors $n^k$ to $\S$ so that
$n^1= \{ n^1_{\phi}, 0, ...,0 \}$ and $n^2= \{0, n^2_{\rho}, 0, ..., 0 \}$.
Then the $\epsilon$-tensor reads
$$
\epsilon_{\phi \rho i_1 ... i_{2p-2}}=n_{\phi}^{1} n^2_{\rho} \epsilon_{i_1 ...
i_{2(p-1)}}
$$
where $\epsilon_{i_1 ... i_{2(p-1)}}$ is the rank $2(p-1)$ Levi-Civita
tensor on $\S$. Due to the orthonormality of the vectors $n^k$, the product of
$\epsilon$-tensors in (\ref{fun1}) becomes the product of their
$2(p-1)$-dimensional analogs
$$
\epsilon_{\phi \rho i_1 ... i_{2p-2}}
\epsilon^{\phi \rho j_1 ... j_{2p-2}}=
\epsilon_{i_1 ... i_{2p-2}}
\epsilon^{j_1 ... j_{2p-2}}
$$
Besides, so far as the extrinsic
curvatures of the surface $\S$ vanish due to the isometry, the Gauss-Codacci
equations \cite{GC} enable one to identify $R^{i_k i_n}_{\ \ j_l j_m}$
on $\S$ with the components of the Riemann tensor of this surface.
Thus, in the limit $a\rightarrow 0$ one obtains the integral
\begin{equation}
\chi[{\cal M}_{\alpha}]= c_p\int_{{\cal M}_
{\alpha}/\Sigma}
{\cal L}_p + 8\pi p c_p(1-\alpha) \int_{\S}^{}{\cal L}_{(p-1)}
\end{equation}
where the first term in r. h. s. is evaluated in the regular points
of ${\cal M}_{\alpha}$ and ${\cal L}_{(p-1)}$ takes the form (\ref{fun})
defined with respect to
the metric on $\S$. Finally, comparing this with (\ref{EulerD}) and using
identity $c_{(p-1)}= 8p \pi  c_p $ one
gets the desired formula for the Euler
number (\ref{EulerD}). We will write this for the general case
when ${\cal M}_{\alpha}$ has several singular surfaces $\Sigma_i$ with the
conical deficits $2\pi(1-\alpha_i)$
\begin{equation}
\chi [{\cal M}_{\alpha}]=
c_p\int_{{\cal M}_{\alpha}/\Sigma}{\cal L}_p +
\sum_{i}(1-\alpha_i) \chi [\S_i]
\label{Eulern}
\end{equation}
As was expected the whole expression does not
depend on the regularization and reproduces (\ref{Euler4}) as a particular
case.
This formula is also valid for a two-dimensional
space when the Euler number is proportional to the integral curvature
and the singular surfaces are the point sets. In
this case (\ref{Eulern}) is a consequence of (\ref{limR}) if
one takes into account that $\chi=1$ for a point.

It is worth mentioning as well that (\ref{Eulern}) reminds a
formula for the Euler characteristic of polygons where each vertex
gives a contribution in $\chi$ determined by the corresponding angular defect
\cite{polygon}.

The case is of special interest
when ${\cal M}_{\alpha}$ possesses a continuous isometry rotation
group in the polar coordinate $\phi$ (Eq. (\ref{8})) and
all the singular surfaces in
(\ref{Eulern}) have equal angles $\alpha_i=\alpha$.
Then, if $\alpha=1$, the space is everywhere smooth.
Otherwise, when $\alpha\neq 1$, ${\cal M}_{\alpha}$ can be
obtained by the following chain of continuous topology preserving
deformations:
${\cal M}_{\alpha=1}\rightarrow\tilde{{\cal M}}_{\alpha=1}\rightarrow
\tilde{{\cal M}}_{\alpha}\rightarrow{\cal M}_{\alpha}$.
Therefore, one can identify the Euler numbers
$\chi[{\cal M}_{\alpha}]=\chi[{\cal M}_{\alpha=1}]$, which results, due to
(\ref{Eulern}), in the interesting formula reducing the number $\chi$
of a manifold ${\cal M}_{\alpha}$ to that of the fixed points set of
its abelian isometry
\begin{equation}
\chi[{\cal M}_{\alpha=1}]=\sum_{i}\chi[\Sigma_i]~~~
\label{sumrule}
\end{equation}
where we made use of the fact that for the given case the volume term
in (\ref{Eulern}) equals $\alpha\chi[{\cal M}_{\alpha=1}]$.
Equation (\ref{sumrule}) can be illustrated for the deformed hyperspheres
$S^d_{\alpha}$ \cite{FM}
with the conical deficits of the polar angle. Thus, the singular
set of $S^2_{\alpha}$ consists of its "north" and "south" poles.
Each of these points has $\chi=1$ and
one gets from (\ref{sumrule}): $\chi[S^2]=1+1=2$. On the other hand,
the singular surface of $S^d_{\alpha}$ ($d\geq3$) is $S^{d-2}$
and from (\ref{sumrule}) the known identity
$\chi[S^d]=\chi[S^{d-2}]$ follows.  Note that equation (\ref{sumrule}) is
valid only for spaces with continuous isometry in $\phi$ and it is violated
for arbitrary kind orbifolds with conical singularities.
\bigskip

{\bf B2. Hirtzebruch signature}
\vskip 0.2cm

We confine the analysis to the four-dimensional
case that is of the most importance in applications.
The Hirtzebruch signature $\tau$ on the smooth spaces
without boundaries is represented by
the integral \cite{Euler}
\begin{equation}
\tau={1 \over 96\pi ^2}\int_{\cal M}R_{\mu\nu\alpha\beta}R^{\mu\nu}_{\ \ \
\gamma\sigma}\epsilon^{\alpha\beta\gamma\sigma}\sqrt{g}d^4x~~~.
\label{Hir}
\end{equation}
Consider this integral on the regularized space ${\tilde {\cal M}}_{\alpha}$
and use equation (\ref{reg-con}) to extract the term giving a
singular contribution to the curvature tensor when regularization
is removed. Due to the Levi-Civita tensor, the only additional surface term
that can appear in (\ref{Hir}) is defined by the quantity
$$
R_{\rho\phi ij}R_{(con)}^{\rho\phi}~_{\ \rho\phi}\epsilon^{ij\rho\phi}
$$
where $R_{\rho\phi ij}$ are regular components of the Riemann tensor
taken on the singular surface $\Sigma$ and $ij$ indices are referred to its
coordinates.
However, taking into account the behavior (\ref{8}) of the metric
near $\Sigma$ one can show that $R_{\rho\phi ij}=0$ and the surface
terms are absent. Therefore, the Hirtzebruch signature on ${\cal M}_{\alpha}$
has the same form as that on the smooth manifolds; it is given by the
integral over the regular region
\begin{equation}
\tau[{\cal M}_{\alpha}]=
{1 \over 96\pi ^2}\int_{{\cal M}_{\alpha}/\Sigma}
R_{\mu\nu\alpha\beta}R^{\mu\nu}_{\ \ \
\gamma\sigma}\epsilon^{\alpha\beta\gamma\sigma}\sqrt{g}d^4x~~~.
\label{Hirsing}
\end{equation}
One can also obtain $\tau[{\cal M}_{\alpha}]$ in higher dimensions and show
that, similar to (\ref{Hirsing}), it is represented by the integral over
region ${\cal M}_{\alpha}/\Sigma$ without extra surface terms.
\bigskip

{\bf C. Lovelock gravity}

Now a natural question arises: can one indicate higher order curvature
polynomials not reducible to topological characteristics but still
having strict meaning on the conical singularities? The answer is positive.
To begin with, let us note that the integral of
$(R^2-4R^2_{\mu\nu}+R^2_{\mu\nu\alpha\beta})$ is the topological invariant
only in four dimensions where it is reduced to a total derivative.
Nevertheless, as one can show with the help of (\ref{R1})-(\ref{Riemann1}),
this integral, having extended to higher dimensions,
will be strictly defined as before and can be represented
in the same form as its topological analog (\ref{Euler4}).

The given integral combination is a particular example of the so-called
Lovelock gravity \cite{Lovelock} and its property holds also for
the general Lovelock gravitational action.
This functional is introduced on a d-dimensional Riemannian manifold as
the following polynomial
\begin{equation}
W_L =
\sum_{p=1}^{k_d} \lambda _p \int {1 \over 2^{2p} p!}
\delta_{[\mu_1\mu_2...\mu_{2p-1}\mu_{2p}]}^{[\nu_1 \nu_2 ... \nu_{2p-1}
\nu_{2p}]} R^{\mu_1 \mu_2}_{\ \ \nu_1 \nu_2} ...R^{\mu_{2p-1} \mu_{2p}}
_{\ \ \nu_{2p-1} \nu_{2p}}\equiv
\sum_{p=1}^{k_d} \lambda _p W_p
\label{Lovelock}
\end{equation}
where $\delta_{[...]}^{[...]}$ is the totally antisymmetrized product
of the Kronecker symbols
and $k_d$ is $(d-2)/2$ (or $(d-1)/2$) for even (odd) dimension $d$.
In the 4-dimensional case
there is only one term $W_1=\frac 12 \int R$ in this functional
and it is reduced to the Einstein action. It was argued \cite{Zumino}
that the gravitational action similar to (\ref{Lovelock}) arises in the
low-energy expansion of string models. Moreover, due to antisymmetrization,
no derivatives higher than second order appear in the equations in the
Lovelock theory \cite{Lovelock} and it turns out to be free of ghosts
when expanding about flat space \cite{Zumino}.

The fact that the Lovelock action is a finite and well-defined functional on
manifolds with conical singularities can be proved along the lines
given for the Euler characteristics.
Indeed, each the integral $W_p$ in $W_L$ can be shown by using
the properties of the Levi-Civita tensor to be a dimensional extension
of the corresponding Euler number $\chi$ (\ref{EulerD}).
Thus, the analysis showing that $W_p$ is finite on ${\cal M}_{\alpha}$ and
independent of the regularization is completely the same as that given for
$\chi$. The important things one should use for this are the
antisymmetricity property and a helpful relation
\begin{equation}
\delta^{[\nu_1...\nu_n]}_{[\mu_1...\mu_n]}=\sum_{k=1}^{n}
(-1)^{k+1}\delta ^{\nu_k}_{\mu_1}\delta^{[\nu_2..\nu_1..\nu_n]}
_{[\mu_2..\mu_k..\mu_n]}~~~.
\end{equation}
After a simple algebra the Lovelock action on ${\cal M}_{\alpha}$
can be represented as the sum of the volume and surface parts
\begin{equation}
W_L[{\cal M}_{\alpha}]= W_L[{\cal M}_{\alpha}/\Sigma] + 2\pi (1-\alpha)
\sum_{p=0}^{k_d-1} \lambda _{p+1} W_{p}[\Sigma]
\label{LovelockC}
\end{equation}
where the first term is the action computed at the regular points
and the second one is a Lovelock's action given on the singular surface.
It should be stressed that integrals $W_{p}[\Sigma]$ are defined completely
in terms of the Riemann tensor on $\Sigma$
\begin{equation}
W_p[\Sigma]={1 \over 2^{2p} p!}\int_{\Sigma}\delta^{[i_1...i_{2p}]}
_{[j_1...j_{2p}]}R^{i_1i_2}_{\ \ \ j_1j_2}...R^{i_{2p-1}i_{2p}}_{\ \ \ j_{2p-1}
j_{2p}}~~~
\label{W_p}
\end{equation}
and $W_0\equiv \int_{\Sigma}$.

Formula (\ref{LovelockC}) can be used to investigate the equations
following from the extrema of $W_L({\cal M}_{\alpha})$.
The variations of this functional at fixed
$\alpha$ result in the
normal Lovelock equations \cite{Lovelock} and the surface ones
\begin{equation}
2\pi(1-\alpha) \left(
\sum_{p=1}^{k_d-1}\lambda_{p+1}
\delta_{i i_1i_2...i_{2p-1}i_{2p}}^{j j_1j_2 ... j_{2p-1}j_{2p}}
R^{i_1 i_2}_{\ \ \ j_1 j_2} ...R^{i_{2p-1} i_{2p}}
_{\ \ \ j_{2p-1} j_{2p}}+\lambda_1\delta^{j}_{i}
\right)=\mu \delta^{j}_{i}~~~
\label{Lsurf}
\end{equation}
where $\mu$ is the density of a matter distributed over $\Sigma$.
The latter equation generalizes relation (\ref{eq3}) between string
tension and polar angle deficit in the Einstein theory.
Remarkably, in the higher dimensional case, an essential feature
comes out: even if $\mu=0$, (\ref{Lsurf}) may have non-trivial solutions
different from these with $\alpha=1$. This means that singular
manifolds ${\cal M}_{\alpha}$ can be extrema in the pure Lovelock
gravity. However, further discussion of this point is outside the aim
of this paper.

\bigskip

{\bf D. Calculus of black hole entropy}
\bigskip

Manifolds with conical singularities naturally appear in the path
integral approach
to gravitational thermodynamics in the presence of the Killing horizons
\cite{Su}-\cite{FM},\cite{c7}.
Let the space-time possess a globally defined time-like Killing vector
$\partial_t$ and be static.
Then the free energy of a field system at temperature
$T=\beta^{-1}$ can be shown to coincide, up to multiplier $\beta$, with an
effective action functional $W(\beta)$ given on an Euclidean section
${\cal M}_{\beta}$
of the corresponding background manifold. The time coordinate $\tau$
of this Euclidean space has to be periodical with the period $\beta$.
In the case of the Killing horizon $\Sigma$,
${\cal M}_{\beta}$ acquires conical singularities
on this surface and can be described near it by the metric (\ref{8}) with
$\alpha={\beta \over \beta_H}$. Here $\beta_H^{-1}$
is the Hawking temperature at which conical singularities vanish and
at which the black-hole thermodynamics is considered. However,
to get the entropy $S$ from the partition function $Z(\beta)$
according to
the standard definition
\begin{equation}
S=\left(-\beta{\partial \over \partial \beta}
+1\right) \ln Z(\beta)|_{\beta=\beta _H}
\label{entropy}
\end{equation}
one should put $\beta$ to be slightly different from $\beta_H$.
In terms of the
effective action equation (\ref{entropy}) can be rewritten as
\begin{equation}
S=\left(\alpha {\partial \over \partial\alpha}-1\right)
W({\cal M}_{\alpha})
\label{entropy1}
\end{equation}
where for the background manifold the previous notation ${\cal M}_{\alpha}$
has been introduced and $\alpha\equiv \beta\beta_H^{-1}$. Several examples
how this formula can be used in the framework of the given
regularization approach follow below.

\bigskip

{\bf D1. Higher-derivative gravity}
\vskip 0.2cm

Consider the following gravitational action being quadratic in the curvature
tensor:
\begin{equation}
W=\int \sqrt{g}d^dx\left(-{1 \over 16\pi G}R + a_1 R^2 + a_2R^{\mu\nu}
R_{\mu\nu} + a_3 R^{\mu\nu\lambda\rho}R_{\mu\nu\lambda\rho}\right)~~~
\label{locpart}
\end{equation}
The first term in (\ref{locpart}) is the standard Einstein action, whereas
the others are usually motivated by necessity to get rid off the one-loop
ultraviolet divergences.

Obviously, a straightforward application of (\ref{entropy1}) to calculate the
black hole entropy in such a theory faces a difficulty
so far as the higher order terms are ill-defined on the conical singularities
and first one should change ${\cal M}_{\alpha}$ by its regular analog
\cite{SS}.
Then the formulas (\ref{R1})-(\ref{Riemann1}) give the following expressions
valid for any dimension $d$:
\begin{equation}
\int_{{\cal M}_{\alpha}}R=\alpha\int_{{\cal M}_{\alpha=1}} R
+4\pi(1-\alpha)\int_{\Sigma}~~~,
\label{1}
\end{equation}
\begin{equation}
\int_{{\cal M}_{\alpha}}R^2=\alpha\int_{{\cal M}_{\alpha=1}} R^2
+8\pi(1-\alpha)\int_{\Sigma}R+O((1-\alpha)^2)~~~,
\label{2}
\end{equation}
\begin{equation}
\int_{{\cal M}_{\alpha}}R^{\mu\nu}R_{\mu\nu}=\alpha\int_{{\cal M}_
{\alpha=1}}
R^{\mu\nu}R_{\mu\nu}
+4\pi(1-\alpha)\int_{\Sigma}R_{\mu\nu}n_i^{\mu}n_i^{\nu}+O((1-\alpha)^2)~~~,
\label{3}
\end{equation}
\begin{equation}
\int_{{\cal M}_{\alpha}}R^{\mu\nu\lambda\rho}R_{\mu\nu\lambda\rho}
=\alpha\int_{{\cal M}_{\alpha=1}} R^{\mu\nu\lambda\rho}R_{\mu\nu\lambda\rho}
+8\pi(1-\alpha)\int_{\Sigma}R_{\mu\nu\lambda\rho}n^{\mu}_in^{\lambda}_i
n^{\nu}_j n^{\rho}_j
+O((1-\alpha)^2)~~~,
\label{4}
\end{equation}
where $n^{\mu}_i$ are two orthonormal vectors
orthogonal to the horizon surface $\Sigma$.
To get (\ref{1})-(\ref{4})
we made use of the fact that ${\cal M}_{\alpha}$ is static and of
the Gauss-Codacci identity on $\Sigma$:
$$
R=R_\Sigma+2R_{\mu\nu}n^{\mu}_in^{\nu}_i-R_{\mu\rho\nu\sigma}
n^{\mu}_in^{\nu}_in^{\rho}_jn^{\sigma}_j
$$
in which the second fundamental forms are absent due to the symmetry.
The first integrals in (\ref{1})-(\ref{4})
are defined on the smooth space at $\alpha=1$,
they are proportional to
$\alpha$ and do not affect the entropy $S$. As for the
terms $O((1-\alpha)^2)$ in (\ref{2})-({\ref{4}), they depend on
the regularization prescription and turn out to be singular
in the limit ${\tilde {\cal M}}_{\alpha} \rightarrow{\cal M}_{\alpha}$, but
they do not contribute to $S$ and the energy of the system at the Hawking
temperature $(\alpha=1)$.
Indeed, from (\ref{entropy1}) and (\ref{1})-(\ref{4}) one obtains for
$S$ the following integral over the horizon $\Sigma$
\begin{equation}
S={1 \over 4G} A_\S -\int_{\Sigma}\left( 8\pi a_1 R+ 4\pi a_2
R_{\mu\nu}n^{\mu}_in^{\nu}_i
+8\pi a_3 R_{\mu\nu\lambda\rho}n^{\mu}_i
n^{\lambda}_in^{\nu}_jn^{\rho}_j\right)~~~.
\label{locentropy}
\end{equation}
Remarkably, this expression differs from the Bekenstein-Hawking entropy
$S={1 \over 4G}A_\S$ in
the Einstein gravity by the contributions depending both on internal and
external geometry of the horizon due to higher order curvature
terms in (\ref{locpart}). It is easy
to see that the effect
of internal geometry of $\Sigma$ is reduced to the integral curvature
of this surface. In four dimensions ($d=4$) this, being a topological
invariant,
is an irrelevant addition to $S$.

It is worth noting that exactly
the same expression can be derived by the Noether charge method suggested
by Wald \cite{Wald}. A difference, between two approaches is that the Wald's
method seems to be more general, but it is defined on the equations of
motion, whereas the above derivation of
(\ref{locentropy}) can be also applied off-shell.
A general proof of their equivalence when taken on-shell
has been given in \cite{Nelson}.
\bigskip

{\bf D2. Lovelock gravity}
\vskip 0.2cm

The expression (\ref{locentropy}) can be generalized to the theory with
the gravitational action being an arbitrary polynomial in the Riemann tensor.
A relevant example is again the Lovelock gravity, where the static black-hole
solutions do exist \cite{Wheeler} and their thermodynamics can be treated
along lines of thermodynamics
in the Einstein gravity \cite{Myers1}. The entropy of a hole
in this case can be inferred from the Lovelock action (\ref{LovelockC})
associated with the free energy
\begin{equation}
S=\left(\alpha {\partial \over \partial \alpha} -1\right)W_L({\cal M}_
{\alpha})|_{\alpha=1}=
- 2\pi\sum_{p=0}^{k_d-1} \lambda _{p+1} W_{p}(\Sigma)~~~.
\label{LovelockS}
\end{equation}
and it turns out to depend only on the internal geometry of $\Sigma$.
Formula (\ref{LovelockS}) has been previously derived in the
Hamiltonian approach in \cite{Jacobson}, whereas arguments based
on the dimensional continuation of the Euler characteristics
have been used for its derivation in \cite{c1}.

\bigskip

{\bf D3. Two-dimensional quantum models}
\vskip 0.2cm

Two-dimensional models of quantum gravity represent a remarkable example
when the one-loop effective action $W$ can be found explicitly.
Thus, in the 2d dilaton gravity $W$ is the combination
\begin{equation}
W=W_0 - {c \over 96\pi}W_{PL}
\label{eff}
\end{equation}
of the classical dilaton action
\begin{equation}
W_0=-\int_{}^{}d^2x \sqrt{g} [ F(\Phi) R+G(\Phi) (\nabla \Phi)^2 + U(\Phi)]
\label{dilaton}
\end{equation}
and the Polyakov-Liouville functional (\ref{PL}) generated by the quantum
effects, $c$ is a constant associated with the central charge.

The contribution of classical action $W_0$ (\ref{dilaton}) to the entropy
can easily be found using equation (\ref{entropy}):
\begin{equation}
S_0=4\pi F(\Phi_h)
\label{entdil}
\end{equation}
where $\Phi_h$ is the value of the dilaton field $\Phi$ at the horizon
which in two dimensions is a point $x_h$. This
expression coincides with that previously obtained in \cite{20}.
As for the quantum correction to $S$, it can be derived using formula
(\ref{PLfin})  that defines $W_{PL}$ on conical singularities and the fact
that $X(\alpha)\simeq (1-\alpha)^2$.
{}From (\ref{PLfin}) one immediately finds
\begin{equation}
S_1={c \over 12} \psi (x_h)~~~.
\end{equation}

The total entropy for the effective action (\ref{eff}) reads
\begin{equation}
S=S_0+S_1=4\pi F(\Phi_h)+{c \over 12}\psi_h~~~.
\label{qentr}
\end{equation}
In conformal gauge one puts $\psi_h=\sigma_h$.
Again this result coincides with that previously obtained
by means of the Wald method \cite{Myers}. We seen that the function $\psi (x)$
is not uniquely defined;
one may add any solution of the homogeneous equation $w(x)$: $\Box w=0$.
The concrete choice of $w(x)$ means the specification of the quantum state
of the system and it can be found from appropriate boundary conditions
\cite{Myers}.
Finally, it is worth noting that the last term in (\ref{qentr})
determines a correction which comes from the conformal anomaly and
for the dilaton holes
it leads to a logarithmic dependence of the entropy on the
mass of the hole \cite{SS}, \cite{Fiola}.

\bigskip\bigskip
\begin{center}
{\bf Acknowledgments}
\end{center}

One of the authors (D.F.) thanks Alexander Popov for a number of useful
remarks. This work was partially supported by the International
Science Foundation, grant RFL000.

\newpage
{\appendix \noindent{\large \bf Appendix }}\\
\def\theequation{A.\arabic{equation}}
\setcounter{equation}0

Here we present some technical details omitted in section 2.
Thus, to compute the curvature on the regularized space ${\tilde {\cal M}}_
{\alpha}$, one can take into account that the metric (\ref{reg}) is of
the form $g_{\mu\nu}=e^{\sigma}\tilde{g}_{\mu\nu}$ and then make use of
the formulas for the curvature tensors of two conformally related manifolds
\ba
&&R[g]=e^{-\sigma} \left( R[\tilde{g}]+{1 \over 2}(d-1) \sigma^{\alpha}_{\
\alpha}
\right)~~~, \nonumber \\
&&R^{\mu}_{\ \nu}[g]=e^{-\sigma} \left( R^{\mu}_{\ \nu}[\tilde{g}]
+{1 \over 4}((d-2) \sigma^{\mu}_{\ \nu}+\delta^{\mu}_{\ \nu} \sigma^{\alpha}_{\
\alpha})
\right)~~~, \nonumber \\
&&R^{\mu\nu}_{\ \ \alpha\beta}[g]=e^{-\sigma} \left( R^{\mu\nu}_{\ \
\alpha\beta}
[\tilde{g}]+ \delta^{[ \mu}_{\ [ \alpha} \sigma^{\nu ]}_{\ \beta ]} \right)
\label{A2}
\ea
where
\begin{equation}
\sigma_{\mu\nu} \equiv -2\tilde{\nabla}_\mu \tilde{\nabla}_\nu \sigma+
\tilde{\nabla}_\mu \sigma \tilde{\nabla}_\nu \sigma -
\frac 12 \tilde{g}_{\mu\nu}
(\tilde{\nabla} \sigma)^2~~~.
\label{A3}
\end{equation}

For metric $\tilde{g}_{\mu\nu}$ (\ref{reg}) we have in the vicinity of
$\rho=0$:
\ba
&&\tilde{R}^{\phi}_{\ \rho \phi \rho}= {1 \over a^2} {u'_x \over 2xu}
+O(a^2)~~~,
\nonumber \\
&&\tilde{R}_{\rho\rho}={1 \over a^2} {u'_x \over 2xu}-h(1-{xu'_x \over 2u})
+O(a^2)~~~, \nonumber \\
&&\tilde{R}^{\phi}_{\ \phi}={1 \over a^2} {u'_x \over 2xu^2}-{h \over u}
+O(a^2)~~~, \nonumber \\
&&\tilde{R}={1 \over a^2} {u'_x \over xu^2}+R_\Sigma -{h \over u}(4-{xu'_x
\over u})
+O(a^2)~~~
\label{A4}
\ea
where $h=\gamma^{ij}h_{ij}$ and we introduced the variable $x={\rho \over a}$;
$R_\Sigma$ is the scalar curvature of $\Sigma$.
Other components of the curvature
tensors do not contain the terms divergent in the limit $a \rightarrow 0$.
As for the tensor $\sigma_{\mu\nu}$ (\ref{A3}), near the point $\rho=0$
its components read
\ba
&&\sigma_{\rho\rho}=-8\sigma_1 +4\sigma_1 {xu'_x \over u} +O(a^2)~~~, \nonumber
\\
&&\sigma^{\phi}_{\ \phi}=-8{\sigma_1 \over u} +O(a^2)~~~,
\nn \\
&&\sigma^{\alpha}_{\ \alpha}=-16\sigma_1({1 \over u} -{1 \over 4}{xu'_x \over
u^2})
+O(a^2)~~~.
\label{A5}
\ea
It is easy to see that in the limit $a \rightarrow 0$ only the
two-dimensional conical part of
the metric $\tilde{g}_{\mu\nu}$ gives singular contributions to
the curvature tensors
whereas the terms in (\ref{A4}) result in regular additions.
Finally, taking into account the form of the volume element
$$
d\tilde{\mu}=a^2 e^{-d\sigma/2} u^{1 \over 2}(1+{1 \over 2}a^2x^2
h)\sqrt{\gamma}
xdxd \phi d^{d-2}\theta
$$
one obtains equations (\ref{curv}).
\bigskip

Consider now the integrals of quadratic curvature combinations
on ${\cal M}_\alpha$. By using (\ref{A3}), (\ref{A4}), (\ref{A5})
in the limit $a \rightarrow 0$ it can be shown that
\begin{eqnarray}
&&\int_{{\cal M}_\alpha}^{}R^2=\alpha\int_{{\cal M}_{\alpha=1}}^{}R^2
+
8\pi(1-\alpha)\int_{\Sigma}^{}R_\Sigma -\left(8I_1-{5 \over 2} I_2\right)
\int_{\Sigma}^{}h + \nn \\
&&\left((d-4)I_2-16(d-1)\left(I_1-{ 1 \over 4}I_2\right)\right)
\int_{\Sigma}^{}
\sigma_1 + {1 \over a^2} I_3 A_{\Sigma}~~~,
\label{AR^2}
\end{eqnarray}
\begin{eqnarray}
&&\int_{{\cal M}_\alpha}^{}R^2_{\mu\nu}=\alpha\int_{{\cal M}_{\alpha=1}}^{}
R^2_{\mu\nu}
+
\left({3 \over 4} I_2-2I_1\right)
\int_{\Sigma}^{}h + \nn \\
&&\left({1 \over 2}(d-4)I_2-4d\left(I_1-{ 1 \over 4}I_2\right)\right)
\int_{\Sigma}^{}
\sigma_1 +
{1 \over 2a^2} I_3 A_{\Sigma}^{}~~~,
\label{ARicci}
\end{eqnarray}
\begin{eqnarray}
&&\int_{{\cal M}_\alpha}^{}R^2_{\mu\nu\alpha\beta}
=\alpha\int_{{\cal M}_{\alpha=1}}^{}R^{2}_{\mu\nu\alpha\beta}
+{1 \over 2} I_2
\int_{\Sigma}^{}h + \nn \\
&&\left((d-4)-16\left(I_1-{ 1 \over 4}I_2\right)\right)\int_{\Sigma}^{}
\sigma_1
+ {1 \over a^2} I_3 A_{\Sigma}^{}
\label{ARiemann}
\end{eqnarray}
where $A_\Sigma=\int_{\Sigma}^{}\sqrt{\gamma}d^{d-2}\theta$ is the area of
the singular surface $\Sigma$.
By means of identities (\ref{A2}) the integrals $\int_{\Sigma}^{}h$  and
$\int_{\Sigma}^{}\sigma_1$ can be written in a coordinate invariant form
in terms of the curvature tensors for the initial metric $g_{\mu\nu}$
(\ref{8}):
\ba
&&\int_{\Sigma}^{}h=\int_{\Sigma}^{}\left({d \over 4}
R_{\mu\rho\nu\lambda}n^{\mu}_in^{\nu}_in^{\lambda}_jn^{\rho}_j-{1 \over 2}
R_{\mu\nu}n^{\mu}_in^{\nu}_i\right) \nn \\
&&8(d-1)\int_{\Sigma}^{}\sigma_1=\int_{\Sigma}^{}(R_\Sigma-R- d
R_{\mu\rho\nu\lambda}n^{\mu}_in^{\nu}_in^{\lambda}_jn^{\rho}_j
+2R_{\mu\nu}n^{\mu}_in^{\nu}_i)~~~.
\label{A11}
\ea
Finally, when using (\ref{A11}), expressions (\ref{AR^2}),
(\ref{ARicci}), (\ref{ARiemann})
take the invariant form of equations
(\ref{R1}), (\ref{Ricci1}), (\ref{Riemann1}).

\end{document}